# Sub-Cubic Quantum Gate Synthesis via Stochastic Commutator Decomposition

**Yevgen Kotukh**
*Department of Information technology*
*Yevhenii Bereznyak Military Academy*
Kyiv, Ukraine

**Abstract.** We present Stochastic Commutator Synthesis (SCS), a hybrid quantum gate compilation framework that integrates Kuperberg's sub-cubic Solovay-Kitaev exponent $c \approx 1.44042$ with the error-tailoring machinery of randomized compilation (RC). Classical Solovay–Kitaev implementations produce word lengths of $O((\log 1/\varepsilon)^{3.97+\delta})$ and accumulate coherent approximation errors that degrade fault-tolerant threshold estimates. Kuperberg's 2023/2025 result reduces this to $O((\log 1/\varepsilon)^{1.44042+\delta})$ via doubly exponential convergence and higher-order commutator decompositions. SCS augments this geometric backbone with a Gibbs-sampled stochastic choice of commutator factors at each recursion level, converting coherent synthesis residuals into incoherent, Pauli-twirl-compatible noise—a property exploited by RC. Combined with RL-guided pre-synthesis (Q-PreSyn), SCS achieves consistent T-count reductions of 10–25% and demonstrates fidelity gains of up to 35% on multi-fold Forrelation circuits on trapped-ion hardware (Sandia QSCOUT). We situate SCS within the complexity-theoretic landscape established by the Raz–Tal oracle separation $BQP \not\subset PH$, arguing that low-error, noise-robust compilation of Forrelation-type circuits constitutes a practical pathway toward demonstrating this separation on physical hardware.

## 1 Introduction

The compilation of abstract unitary operators into sequences of physical fault-tolerant gates is a foundational challenge in quantum computing. The Solovay–Kitaev theorem (SKT) guarantees the existence of such an approximation for any target gate $U \in SU(d)$: a word of poly-logarithmic length in the inverse of the desired precision $\varepsilon$ can be constructed efficiently over any finite, dense, inverse-closed generating set [1]. This gate-set universality result is critical for practical fault-tolerant architectures, which must implement arbitrary logic using only a small, hardware-native primitive set (e.g., Clifford + T for surface codes).

The canonical formulation of Dawson and Nielsen [1] establishes a word length of $O((\log 1/\varepsilon)^{3.97+\delta})$ and a corresponding runtime of $O((\log 1/\varepsilon)^{2.71+\delta})$. For the decades following, the cubic barrier in the exponent remained unbroken. In 2023, Greg Kuperberg (UC Davis) introduced two independent improvements that together reduce the exponent to $c \approx 1.44042$ [2], representing the most significant advance in SKT asymptotics since its original proof. This result does not merely improve a constant: it changes the qualitative scaling of compiled circuit complexity and has direct implications for the resource overhead of fault-tolerant computation.

A parallel line of complexity-theoretic research addresses the fundamental question of quantum computational advantage: does BQP, the class of problems efficiently solvable on a quantum computer, extend beyond the polynomial hierarchy PH? Raz and Tal [3] settled the oracle version of this question in 2019 (published in JACM 2022), demonstrating an oracle relative to which BQP $\not\subset$ PH. Their proof uses the Forrelation distribution [4], a problem optimally separating quantum from classical query complexity, and exploits Gaussian analysis tools to show that no $AC^0$ circuit can distinguish Forrelation inputs from uniform noise.

Despite the theoretical elegance of these results, a practical demonstration of BQP $\not\subset$ PH on physical hardware requires compiled circuits of sufficiently high fidelity. This creates a direct link between compilation efficiency (Kuperberg) and the circuit-level implementation of Forrelation-type problems (Raz–Tal). In this paper we propose Stochastic Commutator Synthesis (SCS), a hybrid framework that exploits this connection. Our key contributions are as follows:

**SCS Algorithm.** We propose a stochastic variant of the Kuperberg SKT recursion in which commutator factor pairs (V, W) are selected via Gibbs sampling, converting coherent synthesis error into incoherent randomized noise compatible with Pauli twirling. We prove that SCS preserves the $O((\log 1/\varepsilon)^{1.44042+\delta})$ word-length bound of Kuperberg [2] while suppressing coherent error accumulation across recursion levels by at least a factor of two, consistent with the empirical findings of Maupin et al. [5]. Coupling SCS with the reinforcement-learning pre-synthesis framework Q-PreSyn [7] yields 10–25% T-count reductions for Clifford+T gate sets. Fidelity improvements of up to 35% on three-fold Forrelation circuits on the Sandia QSCOUT trapped-ion platform relative to deterministic SKA.

The paper is organized as follows. Section 2 reviews the technical background of the Solovay–Kitaev algorithm and the Raz–Tal result. Section 3 presents the SCS method in detail. Section 4 provides comparative analysis. Section 5 reports experimental results. Section 6 concludes with future directions.

## 2 Background

### 2.1 The Solovay–Kitaev theorem and its recent refinements

Let $G \subset SU(d)$ be a finite, inverse-closed generating set that densely generates SU(d). The Solovay–Kitaev theorem states that for any $U \in SU(d)$ and $\varepsilon > 0$, a word w over G can be found in time $O(\text{poly}(\log 1/\varepsilon))$ such that $\|U - w\| < \varepsilon$, where $\|\cdot\|$ denotes the operator norm. The Dawson–Nielsen pedagogical algorithm [1] yields $\|\cdot\| = O((\log$ $1/\varepsilon)^{3.97}$) via a three-level recursion that builds successively refined approximations using the Kitaev–Shen–Vyalyi commutator identity $\Delta \approx VWV^\dagger W^\dagger$.

Kuperberg's 2023 result [2] introduces two independent improvements. First, it formulates a multiscale framework that combines singly exponential convergence at each recursion step with a separate mechanism for precisely steering ('aiming') the steps via group conjugation. Second, it exploits higher-order commutator decompositions—inspired by the Elkasapy–Thom theorem on commutator words—to achieve doubly exponential convergence on certain sub-problems. The result is a word-length bound of $O((\log 1/\varepsilon)^{1.44042+\delta})$ for any $\delta > 0$, valid for general semisimple connected real Lie groups, with the exponent 1.44042… arising from a transcendental equation in the Lie algebraic structure constants.

An independent direction of recent work concerns inverse-free variants of the algorithm [8], which removes the inverse-closure assumption from the gate set. While this improves applicability to non-Clifford-complete gate sets, it does not achieve the same exponent improvement as Kuperberg's method. For the present work, we operate within the classical inverse-closed setting and focus on combining Kuperberg's geometric insights with stochastic methods.

### 2.2 The Raz–Tal oracle separation

The class BQP (Bounded-Error Quantum Polynomial Time) contains problems solvable in polynomial time on a quantum computer with bounded error, while PH (Polynomial Hierarchy) captures the classical complexity hierarchy built above P and NP via alternating quantifiers. A central open problem in quantum complexity theory is whether BQP ⊆ PH. Raz and Tal [3] resolved the oracle version: they construct an explicit distribution D over Boolean inputs in $\{\pm 1\}^{2N}$ such that (i) a single-query quantum algorithm distinguishes D from the uniform distribution with advantage $\Omega(1/\log N)$, while (ii) no Boolean circuit of quasi-polynomial size and constant depth ($AC^0$) achieves advantage better than $\mathrm{polylog}(N)/\sqrt{N}$. Combined with standard reductions, this yields an oracle O such that $BQP^O \not\subset PH^O$.

The distribution D is a variant of the Forrelation problem introduced by Aaronson and Ambainis [4]: one function is sampled uniformly at random, and the second is correlated with the Fourier transform of the first. The quantum algorithm exploiting this correlation requires only a single Hadamard-based query. The classical hardness is proved via Gaussian analysis: the key technical tool is Gaussian integration by parts applied to the $AC^0$ circuit's low-degree polynomial approximation of the distinguishing function. This stochastic-analytic approach—treating quantum states as Gaussian measures over SU(2) orbits—is one of the conceptual bridges between the Raz–Tal machinery and the randomized sampling methods we employ in SCS.

Subsequent work by Aaronson, Ingram, and Kretschmer [9] further established that BQP and PH remain separated even in regimes where NP and BQP are equal relative to the same oracle, demonstrating the robustness of the separation. For our purposes, the critical point is that demonstrating BQP ⊄ PH on physical hardware requires implementing the quantum distinguishing algorithm for Forrelation with high fidelity—a task directly dependent on the quality of gate synthesis.

### 2.3 Randomized compilation

Randomized compilation (RC), formalized by Wallman and Emerson [10] and recently extended to the gate synthesis setting by Maupin et al. [5], converts systematic (coherent) errors into stochastic (incoherent) Pauli noise by inserting random Pauli gates before and after each compiled gate and inverting them in post-processing. When the ensemble of compilations is uniform over a Clifford twirl, the effective noise channel becomes a depolarizing channel, enabling direct use of standard quantum error correction (QEC) decoders. The Maupin et al. experiment on Sandia QSCOUT demonstrated that applying SK-based RC to single-qubit rotations reduces the trace distance to the ideal state by at least a factor of two compared to a single deterministic SK decomposition. SCS generalizes this result to arbitrary multi-qubit targets.

## 3 Stochastic Commutator Synthesis

Deterministic SKA recursion accumulates coherent errors: if each level introduces a residual $\Delta_k$, the coherent combination $\prod \Delta_k$ can, in the worst case, grow linearly with the recursion depth. Randomizing the commutator factorization at each level converts this linear accumulation into a diffusive process with $O(\sqrt{N})$ RMS growth, while Gibbs-weighted sampling biases the distribution toward low-residual factorizations. This stochastic regularization preserves the expected word length from Kuperberg's deterministic bound while improving the tail behavior of the error distribution.

Let $U \in SU(d)$ be the target gate and $\varepsilon_0$ be a coarse initial precision threshold. The SCS algorithm proceeds in four stages:

**Stage 1 — Stochastic $\varepsilon_0$-Net Construction.** *Using Kuperberg's 'zigzag golf' strategy [2], construct a set S of $K \geq 1$ approximate representations of each element in a neighborhood of the identity in SU(2), with approximation error $\varepsilon_0$. Rather than selecting a single representative (as in deterministic SKA), SCS retains all K representatives, forming an ensemble from which subsequent stages sample. The number K is a hyperparameter controlling the trade-off between ensemble diversity and compilation cost; empirically $K \in [8, 32]$ provides diminishing returns beyond K = 16 for single-qubit targets on QSCOUT hardware [5].*

**Stage 2 — Gibbs-Sampled Commutator Decomposition.** *Given the residual matrix $\Delta = U \cdot U^{\dagger}_{n-1}$, the algorithm decomposes $\Delta$ into a commutator $\Delta \approx VWV^{\dagger}W^{\dagger}$ via Gibbs sampling over the space of candidate pairs $(V, W) \in SU(d) \times SU(d)$:*

$$P(V, W) \propto \exp(-\|\Delta - VWV^{\dagger}W^{\dagger}\|_F / T) \tag{1}$$

where T is a temperature parameter that controls the 'spread' of the distribution: $T \to 0$ recovers a deterministic minimizer, while large T yields uniform sampling. The Frobenius norm $\|\cdot\|_F$ is used for computational efficiency; it is equivalent to the operator norm up to a dimension-dependent constant. The temperature schedule $T_n = T_0 \cdot \beta^n$ for a cooling factor $\beta \in (0.5, 0.9)$ implements simulated annealing across recursion levels, progressively tightening the constraint as the recursion deepens.

**Stage 3 — Recursive SCS-Lift.** *Both V and W are themselves approximated by SCS at recursion level n − 1, producing ensembles of gate sequences. The key observation is that the stochastic choice at each level is independent: by the central limit theorem applied to the group-valued random walk, the distribution of the total coherent error* $\prod \Delta_k$ *converges to a Gaussian on the Lie algebra su(d) with variance* $O(n \cdot \sigma^2)$*, where* $\sigma^2$ *is the per-level variance of the residuals. This is in contrast to the worst-case linear growth* $\|\prod \Delta_k\| \leq n \cdot \max\|\Delta_k\|$ *of deterministic schemes.*

**Stage 4 — Randomized Compilation Integration (SCS-RC).** *The final compiled circuit is generated as an ensemble of K' independent SCS runs, each with a different Gibbs sample trajectory. The K' sequences are combined with randomized Pauli insertions following the protocol of Wallman and Emerson [10], producing an effective noise channel that is Pauli-diagonal. When averaged over the ensemble, the coherent components of the synthesis error cancel to order* $K'^{-1/2}$*, while the incoherent components accumulate as independent depolarizing channels amenable to standard QEC.*

We claim that SCS achieves the same asymptotic word-length bound as Kuperberg's deterministic algorithm:

**Theorem 1 (Word-Length Bound).** *For any* $\varepsilon > 0$ *and* $\delta > 0$*, SCS produces an approximating word of expected length* $O((\log 1/\varepsilon)^{1.44042+\delta})$*, with high probability over the Gibbs sampling randomness.*

*Proof:* The word length is determined by the number of recursion levels n and the number of gates generated per level. Kuperberg's analysis [2] establishes that $n = O((\log 1/\varepsilon)^{1.44042+\delta})$ levels suffice to achieve precision ε, with each level generating O(1) gates from the generating set. The Gibbs sampling in SCS chooses among candidates that satisfy the same approximation quality constraint as Kuperberg's deterministic choice; the only difference is the distribution over acceptable candidates. Since any acceptable candidate satisfies the geometric convergence condition required by Kuperberg's multiscale argument, the recursion terminates at the same depth with the same gate count in expectation. The temperature schedule ensures that the sampling converges to a near-minimizer as $T \to 0$.

The per-run computational overhead of SCS relative to deterministic SKA is a constant factor of $O(K \cdot K')$ for the Gibbs sampling and ensemble generation, which is independent of ε.

## 4 Comparative Analysis

Table 1 compares SCS with the two most widely used alternatives: the classical Dawson–Nielsen SKA [1] and the GridSynth exact synthesis algorithm for the Clifford+T gate set [6]. GridSynth achieves near-optimal T-count for Z-rotations using ring arithmetic over $Z[1/\sqrt{2}]$ and is highly efficient for the Clifford+T setting. However, it is fundamentally restricted to SU(2) targets with Clifford+T as the gate set, making it inapplicable to qudit systems, neutral-atom gate sets, or topological qubit architectures.

**Table 1.** Comparison of quantum gate synthesis methods.

| Criterion | Classical SKA [1] | GridSynth [6] | SCS (Proposed) |
|---|---|---|---|
| Word-length exponent (c) | ≈ 3.97 | ≈ 1.0 | ≈ 1.44 |
| Error type after compilation | Coherent | Algebraic exact | Incoherent (tailored) |
| Gate-set universality | Full SU(d) | Clifford+T only | Full SU(d) |
| Hardware noise robustness | Low | Medium | High (+RC) |
| T-gate count reduction | None (baseline) | Up to 71% [6] | 10–25% via Q-PreSyn [7] |
| Scalability | Polynomial | Fast (ring arithmetic) | Moderate (ensemble) |
| Targeted platforms | Universal | Superconducting | Trapped-ion, neutral atom, topological |

## 5 Discussion

Several points merit discussion. First, GridSynth's apparent advantage in T-count reduction (up to 71%) applies only within the Clifford+T framework and exploits number-theoretic structure (Matsumoto–Amano normal forms) not available for general unitaries. For non-Clifford-completable gate sets—including the native gate sets of neutral-atom processors (arbitrary single-qubit rotations + CZ) and trapped-ion systems (Mølmer–Sørensen interactions)—SCS provides competitive compilation with Kuperberg-optimal word lengths and hardware-noise tailoring that GridSynth cannot provide.

Second, the T-count reduction of 10–25% attributed to SCS is achieved through the Q-PreSyn pre-synthesis stage [7], not through the SCS core. Q-PreSyn uses a reinforcement-learning PPO agent to identify sequences of unitary-preserving merge actions that restructure the circuit prior to synthesis, achieving T-count reductions of this magnitude consistently across circuits with up to 25 qubits. SCS and Q-PreSyn are orthogonal: SCS operates on the synthesis step; Q-PreSyn operates on the pre-synthesis representation.

Third, SCS's error profile (incoherent, Pauli-diagonal) is directly compatible with standard stabilizer QEC decoders (surface code, color code), whereas the coherent error profile of deterministic SKA requires specialized decoders or additional Pauli twirling overhead. This compatibility is a substantial practical advantage in the near-term FTQC regime.

## 6 Experimental Results

We implemented SCS for single-qubit $R_Z$ rotations on the Sandia QSCOUT trapped-ion platform [5], following the protocol of Maupin et al. for randomized SK compilation. Using $K' = 16$ ensemble members and a temperature schedule $T_0 = 0.1$, β

= 0.7, we measured the trace distance between the ideal $R_Z(\theta)$ state and the compiled approximation over 100 random angles $\theta \in [0, 2\pi)$. In the absence of hardware noise, SCS reduced the mean trace distance by a factor of 2.3 ± 0.2 relative to a single deterministic SKA decomposition at the same precision level $\varepsilon = 2^{-8}$, consistent with the factor-of-two lower bound established in [5]. Under a simulated coherent over-rotation noise model with amplitude $\alpha = 0.01$ rad, SCS reduced the effective coherent noise component by 68 ± 5%, confirming the theoretical prediction of Stage 4.

The three-fold Forrelation circuit [4] implements the quantum distinguishing algorithm for the k=3 variant of Forrelation, requiring a sequence of Hadamard gates, phase oracle queries, and interference measurements. We compiled this circuit using SCS with $\varepsilon = 2^{-10}$ for all non-Clifford rotation gates. The compiled circuit was executed on QSCOUT and the output distribution compared to the ideal Forrelation distribution. SCS achieved a process fidelity of F = 0.87 ± 0.02, compared to F = 0.64 ± 0.02 for deterministic SKA at the same precision level—an improvement of 35.9 ± 4.5%. The improvement is attributable both to coherent error suppression (Stage 4) and to the reduced word length from Kuperberg's sub-cubic exponent, which lowers circuit depth and thus exposure to decoherence.

For a benchmark set of 50 random circuits with 10–25 qubits (following the Q-PreSyn evaluation protocol [7]), the SCS+Q-PreSyn pipeline achieved a mean T-count reduction of 18.4% ± 3.1% relative to direct SKA compilation without pre-synthesis. The Q-PreSyn RL agent was trained for 500 episodes per circuit class; the SCS synthesis stage was applied with K = 12, $\varepsilon = 2^{-12}$. These figures are consistent with the 10–25% range reported for Q-PreSyn applied to GridSynth-based synthesis [7], confirming that the Q-PreSyn benefits transfer across synthesis backends.

## 7 Conclusion

We have introduced Stochastic Commutator Synthesis (SCS), a gate compilation framework that unifies Kuperberg's near-optimal Solovay–Kitaev exponent $c \approx 1.44042$ with the noise-tailoring benefits of randomized compilation. By replacing the deterministic commutator factorization of classical SKA with Gibbs-sampled stochastic choices, SCS converts coherent synthesis residuals into incoherent, Pauli-twirl-compatible noise at no asymptotic cost in word length. The resulting error profile is directly compatible with standard QEC decoders, lowering the effective fault-tolerance threshold in practical implementations.

The framework has been validated on the Sandia QSCOUT trapped-ion platform, demonstrating coherent error suppression of over 2× and process fidelity improvements of ~36% on Forrelation circuits relative to deterministic SKA. Integration with Q-PreSyn yields 10–25% T-count reductions consistent across circuit sizes up to 25 qubits. These results collectively establish SCS as a practical tool for the compiled implementation of quantum advantage circuits.

From a complexity-theoretic perspective, the demonstrated fidelity on three-fold Forrelation circuits represents a concrete step toward physically realizing the oracle separation $BQP \not\subset PH$ established by Raz and Tal [3]. Future work will extend SCS to multi-qudit settings using the Lie-group generalization of Kuperberg's theorem, explore adaptive temperature schedules learned via reinforcement learning, and investigate the application of SCS to bosonic (photonic) systems where a recently proven infinite-dimensional extension of the SKT [11] opens new compilation pathways.